\documentclass[twocolumn,showpacs,preprintnumbers,amsmath,amssymb,showkeys,nofootinbib]{revtex4}
\usepackage{amsmath,amssymb}
\usepackage{graphicx}

\usepackage[english]{babel}
\newcommand{\be}{\begin{equation}}
\newcommand{\ee}{\end{equation}}

\begin{document}
\title{Chimeras in networks with purely local coupling}% Force line breaks with \\

\author{Carlo R. Laing}
\email{c.r.laing@massey.ac.nz}
\affiliation{ Institute of Natural and Mathematical Sciences, 
Massey University, Private Bag 102-904 NSMC, Auckland, New Zealand. \\
phone: +64-9-414 0800 extn. 43512
fax: +64-9-443 9790 }

\date{\today}

\begin{abstract}
Chimera states in spatially extended networks of oscillators have some oscillators synchronised while the remainder
are asynchronous. These states have primarily been studied in networks with nonlocal coupling, and more recently
in networks with global coupling. Here we present three networks with only local coupling
(diffusive, to nearest neighbours) which are numerically found to support chimera states.
One of the networks is analysed using a self-consistency argument in the continuum limit, and  
this is used to find the
boundaries of existence of a chimera state in parameter space.

\end{abstract}

\maketitle

%\section{Introduction}

Chimera states, in which a symmetric network of identical oscillators splits into two regions, one of coherent
oscillators and one of incoherent, have been studied intensively over the past decade~\cite{kurbat02,abrstr06,panabr15}.
Spatial networks on which they have been studied include a 
one-dimensional ring~\cite{abrstr06,ome13,lai09A,lai09B,abrstr04,wolome11}, a square domain
without periodic boundary conditions~\cite{marlai10,gust13,shikur04}, a torus~\cite{panabr13,omewol12} and 
a sphere~\cite{panabr15a}. They have also been observed recently in a number of experimental 
settings~\cite{tinnko12,hagmur12,marthu13,nkotin13,wickis14}.
Early investigations considered networks with nonlocal coupling (i.e.~neither all-to-all
with uniform strength, nor local coupling, via diffusion, for example) as chimeras were first reported
in nonlocally coupled systems~\cite{kurbat02,kurshi03}. Nonlocal coupling was at first thought to be essential
for the existence of chimeras, however, more recent results show that chimeras can occur in systems with
purely global coupling~\cite{schsch14,setsen14,yelpik14}.

Here we consider the opposite limit and address the question as to whether chimera states can exist in 
spatial networks with purely local coupling. We present three such networks in which this does occur.
The idea behind the creation of the networks is straightforward and can be found in the early
papers~\cite{kurshi03,shikur04}. Consider a general reaction-diffusion equation
on a one-dimension spatial domain $\Omega$ with only local interactions via diffusion in one variable:
\begin{align}
   \frac{\partial u}{\partial t} & =f(u)+v \label{eq:du} \\
   \epsilon\frac{\partial v}{\partial t} & = g(u)-v+\frac{\partial^2 v}{\partial x^2} \label{eq:dv}
\end{align}
Setting $\epsilon=0$ in~\eqref{eq:dv} one has 
\be
   \left(1-\frac{\partial^2}{\partial x^2}\right)v=g(u) \label{eq:diff}
\ee
and if $h(x)$ is the Green's function associated with $\left(1-\frac{\partial^2}{\partial x^2}\right)$ 
on $\Omega$ then
we can write~\eqref{eq:diff} as
\be
  v(x)=\int_\Omega h(x-y)g(u(y))\ dy
\ee
and substituting this into~\eqref{eq:du} we obtain a closed nonlocal equation for $u$.
We will implement the network analogue of~\eqref{eq:du}-\eqref{eq:dv} but with $\epsilon$ small and nonzero
in the expectation that the behaviour of interest when $\epsilon=0$ persists for $\epsilon>0$.

\begin{figure}[t]
\includegraphics[scale=0.45]{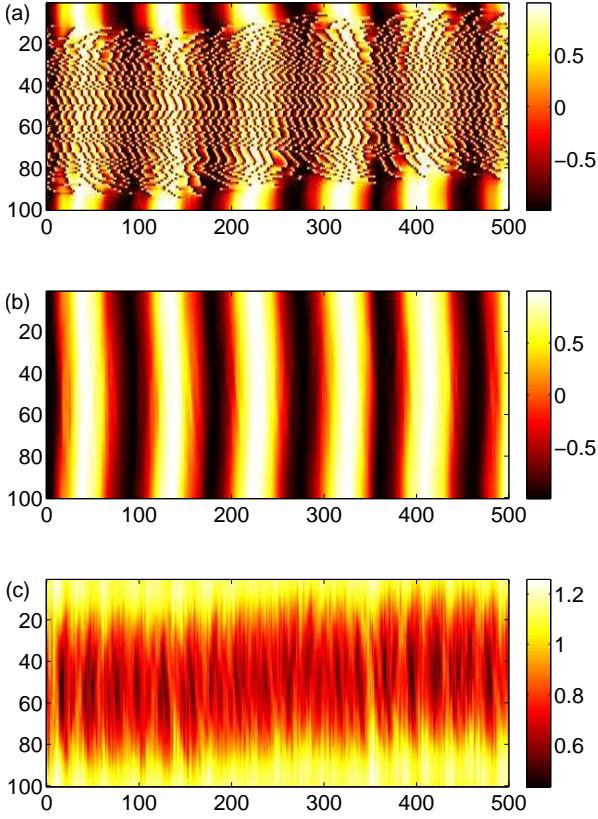}
\caption{(Color online) Chimera solution of the system~\eqref{eq:dth}-\eqref{eq:dz}.
(a): $\sin{\theta_j}$; (b): $\sin{(\arg{(z_j)})}$; (c): $|z_j|$.
Parameters: $\omega_0=1,\sigma=0.01,\epsilon=0.2,A=1.5,L=2\pi,N=100,\beta=0.1$.}
\label{fig:exam}
\end{figure}

The first model we consider consists of $N$  oscillators, equally-spaced on a domain of length $L$, with
periodic boundary conditions. The state of oscillator $j$ is described by two variables:
$\theta_j\in[0,2\pi)$ and $z_j\in\mathbb{C}$. (A complex variable is used to simplify presentation;
we could equally well use two real variables.) The governing equations are
\begin{align}
   \frac{d\theta_j}{dt} & = \omega_j-\mbox{Re}\left(z_je^{-i\theta_j}\right) \label{eq:dth} \\
   \epsilon\frac{dz_j}{dt} & = Ae^{i(\theta_j+\beta)}-z_j+\frac{z_{j+1}-2z_j+z_{j-1}}{(\Delta x)^2} \label{eq:dz}
\end{align}
for $j=1,2\dots N$, where $\Delta x=L/N$ and $A,\beta$ and $\epsilon$ are all constants. The $\omega_j$ are randomly
chosen from a Lorentzian
distribution with half-width-at-half-maximum $\sigma$ centred at $\omega_0$, namely
\be
   g(\omega)=\frac{\sigma/\pi}{(\omega-\omega_0)^2+\sigma^2} \label{eq:lor}
\ee
An example
of the system's dynamics are shown in Fig.~\ref{fig:exam}. The domain clearly splits into two regions,
one showing coherent behaviour of the phases and the other, incoherent. This behaviour has been replicated in networks
of up to $N=1000$, so is not a small-$N$ effect.

To understand the relationship between~\eqref{eq:dth}-\eqref{eq:dz} and previously studied models we 
set $\epsilon=0$ in~\eqref{eq:dz}. If $z_j$ is the $j$th entry of the vector ${\bf z}\in\mathbb{C}^N$ 
and similarly for $\theta_j$ we can write~\eqref{eq:dz} as
\be
   (I-D){\bf z}  =Ae^{i{(\boldsymbol{ \theta}+\beta)}} 
\ee
where $I$ is the $N\times N$ identity matrix and $D$ is the matrix representation of the
classical second difference operator on $N$ points with periodic boundary conditions.
Defining $G=(I-D)^{-1}$ we have
\be
   z_j  = A\sum_{k=1}^NG_{jk}e^{i{(\theta_k+\beta)}} \label{eq:z}\\
\ee
where $G_{jk}$ is the $jk$th element of $G$, and substituting~\eqref{eq:z} into~\eqref{eq:dth}  we obtain
\be
   \frac{d\theta_j}{dt} = \omega_j-A\sum_{k=1}^N G_{jk}\cos{(\theta_j-\theta_k-\beta)} \label{eq:dthe0}
\ee
Since $I-D$ is circulant, so is $G$, i.e.~$G_{ij}$ is a function of only $|i-j|$, and~\eqref{eq:dthe0}
is thus of the same form as studied by a number of others~\cite{ome13,lai09A,abrstr04,kurbat02,lai09B,abrstr06,wolome11}.
An important property of~\eqref{eq:dthe0} is that it is invariant with respect to a uniform shift of all
phases: $\theta_j\mapsto\theta_j+\gamma$ for all $j$ where $\gamma$ is some constant. This implies that~\eqref{eq:dthe0}
can be studied, without loss of generality, 
in a rotating coordinate frame where $\omega_0=0$, i.e.~the actual value of $\omega_0$ in~\eqref{eq:dthe0}
is irrelevant. This is not the case for~\eqref{eq:dth}-\eqref{eq:dz} when $\epsilon\neq 0$ 
(although~\eqref{eq:dth}-\eqref{eq:dz} is invariant under the simultaneous shift: $\theta_j\mapsto\theta_j+\gamma$
and $z_j\mapsto z_je^{i\gamma}$ for all $j$).

\begin{figure}[t]
\includegraphics[scale=0.45]{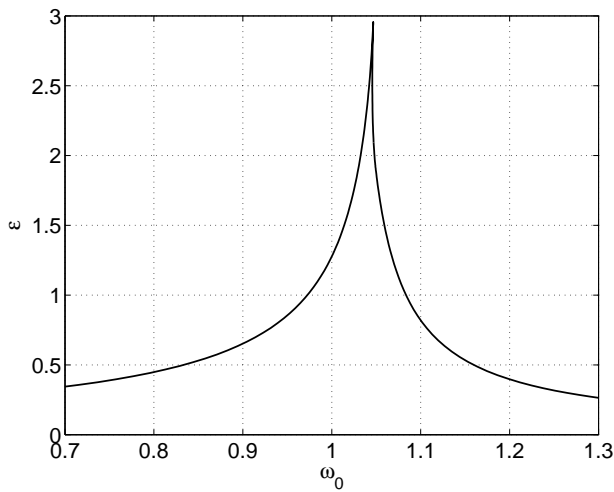}
\caption{Saddle-node bifurcation of solutions of~\eqref{eq:self}. Chimera solutions of~\eqref{eq:dth}-\eqref{eq:dz}
are stable below the curve.
Parameters: $A=1.5,L=2\pi,\beta=0.1,\sigma=0.01$.}
\label{fig:epom}
\end{figure}

To analyse the chimera seen in~\eqref{eq:dth}-\eqref{eq:dz} we use a self-consistency argument similar to that
in~\cite{abrstr06,kurbat02,lai09A}. We first move to a rotating coordinate frame,
letting $\phi_j\equiv \theta_j-\Omega t$ and $y\equiv z_je^{-i\Omega t}$, where $\Omega$ is to be determined,
 and then take the limit $N\rightarrow\infty$,
to obtain
\begin{align}
   \frac{\partial\phi}{\partial t} & = \omega-\mbox{Re}\left(ye^{-i\phi}\right)-\Omega \label{eq:dphi} \\
   \epsilon\frac{\partial y}{\partial t} & = Ae^{i(\phi+\beta)}-y+\frac{\partial^2 y}{\partial x^2}-i\epsilon\Omega y \label{eq:dy}
\end{align}
We now search for solutions of~\eqref{eq:dphi}-\eqref{eq:dy} for which $y$ is stationary, i.e.~just a function of
space. We let such a solution be $y=R(x)e^{i\Theta(x)}$. Since $y$ is constant we can  use~\eqref{eq:dphi}
to determine the dynamics of
$\phi$ for any $y$ and $\omega$: if $|R|>|\omega-\Omega|$ then $\phi$ will tend to a stable fixed point of~\eqref{eq:dphi}
whereas if $|R|<|\omega-\Omega|$ then $\phi$ will drift monotonically. To obtain a stationary solution of~\eqref{eq:dy}
we replace $e^{i\phi}$ by its expected value, calculated using the density of $\phi$, which is inversely proportional
to its velocity (given by~\eqref{eq:dphi}). So (keeping mind that $\omega$ is random variable) we need to solve
\be
  0=Ae^{i\beta}\int_{-\infty}^\infty\int_0^{2\pi}e^{i\phi}p(\phi|\omega)g(\omega)d\phi\ d\omega-y+\frac{\partial^2 y}{\partial x^2}-i\epsilon\Omega y
\label{eq:int}
\ee
where the density of $\phi$ given $\omega$ is
\be
  p(\phi|\omega)=\frac{\sqrt{(\omega-\Omega)^2-R^2}}{2\pi|\omega-\Omega-R\cos{(\Theta-\phi)}|}
\ee
and $g(\omega)$ is given by~\eqref{eq:lor}.
Evaluating the integrals in~\eqref{eq:int} we obtain
\begin{multline}
  \frac{Ae^{i(\Theta+\beta)}}{R}\left[\omega_0+i\sigma-\Omega-\sqrt{(\omega_0+i\sigma-\Omega)^2-R^2}\right] \\ 
  -\left(1+i\epsilon\Omega-\frac{\partial^2}{\partial x^2}\right)Re^{i\Theta}=0 \label{eq:self}
\end{multline}
We determine $R,\Theta$ and $\Omega$ by simultaneously solving~\eqref{eq:self} and the scalar equation $\Theta(0)=0$;
this equation amounts to choosing the origin of the rotating coordinate frame. Following solutions of~\eqref{eq:self}
as $\omega_0$ and $\epsilon$ are varied we find that two solutions are destroyed in a saddle-node bifurcation
on the curve shown in Fig.~\ref{fig:epom}. Although our self-consistency argument gives no information about
the stability of solutions (unlike the continuum theory in~\cite{lai09B,ome13,wolome11}) quasistatic sweeps through
parameter space indicate that the curve in Fig.~\ref{fig:epom} does indeed mark the boundary of stable chimeras
in the system~\eqref{eq:dth}-\eqref{eq:dz}. If $\epsilon$ is increased past the boundary in Fig.~\ref{fig:epom}
when $\omega_0$ is to the left of the cusp (at $\omega_0\approx 1.04$) the system~\eqref{eq:dth}-\eqref{eq:dz}
moves to the almost synchronous state, whereas to the right of the cusp the system moves to a spatially-disordered
state, and the almost synchronous state seems unstable here.

\begin{figure}
\includegraphics[scale=0.45]{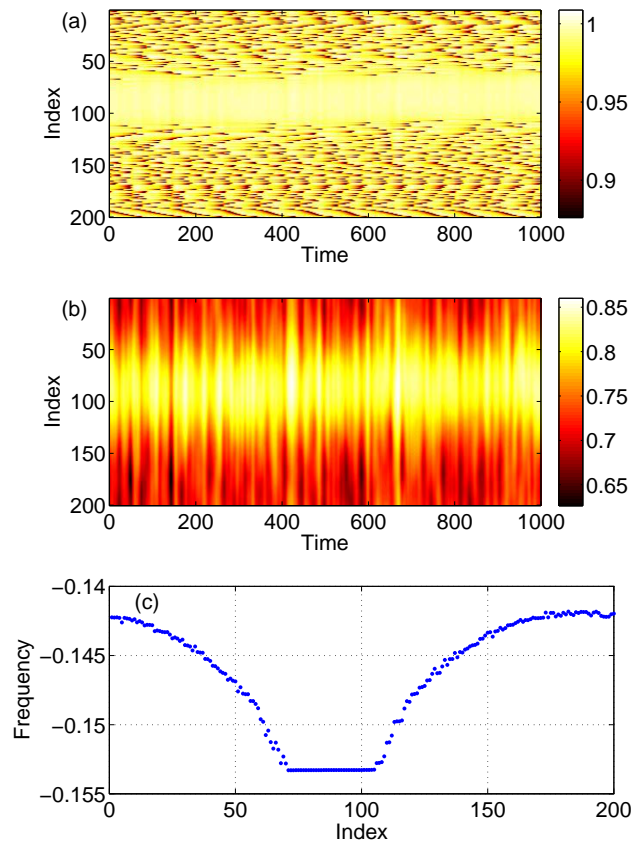}
\caption{(Color online) Chimera solution of the system~\eqref{eq:CGLa}-\eqref{eq:CGLb}.
(a): $|A_j|$; (b): $|Z_j|$; (c): average rotation frequency of the $A_j$ over a simulation of duration $2000$
time units.
Parameters: $\omega_0=0,\epsilon=0.01,a=-1,b=1,K=0.1,N=200$.}
\label{fig:CGL}
\end{figure}

As a second example we consider a network of Stuart-Landau oscillators, each of which can be thought of
as the normal form of a supercritical Hopf bifurcation, with purely local 
coupling through a second complex variable. Using Stuart-Landau oscillators as opposed to the phase oscillators
above introduces an amplitude variable to the oscillator dynamics. 
As above we have $N$ oscillators equally-spaced on a domain
of length $1$ with periodic boundary conditions. The equations are
\begin{align}
   \frac{dA_j}{dt} & = (1+i\omega_0)A_j-(1+ib)|A_j|^2A_j \nonumber \\
    & +K(1+ia)(Z_j-A_j) \label{eq:CGLa} \\
   \epsilon\frac{dZ_j}{dt} & = A_j-Z_j+\frac{Z_{j+1}-2Z_j+Z_{j-1}}{16(\Delta x)^2} \label{eq:CGLb}
\end{align}
for $j=1,2\dots N$ where $A_j,Z_j\in\mathbb{C}$ and $\omega_0,a,b,K$ and $\epsilon$ are real parameters and $\Delta x=1/N$.
Note that the oscillators are identical. A chimera state for this system is shown in Fig.~\ref{fig:CGL}.
Note that setting $\epsilon=0$ in~\eqref{eq:CGLb} and then taking the limit $N\rightarrow\infty$ we obtain the
nonlocally coupled complex
Ginzburg-Landau equation for just the variable $A$, as 
studied by~\cite{kurbat02}. Then assuming that $K$ is small one finds a scale 
separation between the amplitude and phase dynamics of $A$ and upon setting $|A|=1$ the phase dynamics
can be written in a nonlocally coupled form~\cite{kurbat02,panabr15}. 

As a third model we consider a heterogeneous network of oscillators, each described by an angular variable 
and a real variable. The angular variables have the form of Winfree oscillators~\cite{pazmon13,aristr01,win67}.
The model is
\begin{align}
   \frac{d\theta_j}{dt} & = \omega_j+\kappa Q(\theta_j)u_j \label{eq:dwina} \\
   \epsilon\frac{du_j}{dt} & = P_n(\theta_j)-u_j+\frac{u_{j+1}-2u_j+u_{j-1}}{(\Delta x)^2}  \label{eq:dwinb}
\end{align}
for $j=1,2\dots N$ where $Q(\theta)=\sin{\beta}-\sin{(\theta_j+\beta)}$, $\kappa,\beta$ and $\epsilon$ are parameters, $P_n(\theta)=a_n(1+\cos{\theta})^n$ where
$n\geq 1$ is an integer and $a_n=2^n(n!)^2/(2n)!$ (so that $\int_0^{2\pi}P_n(\theta)d\theta=2\pi$) and
$\Delta x=L/N$. The $\omega_j$ are randomly chosen from a normal distribution with mean $\omega_0$ and
standard deviation $\sigma$. $Q(\theta)$ is the phase response curve of the oscillator and can be measured
experimentally for a neuron, for example~\cite{schpri11}.

\begin{figure}
\includegraphics[scale=0.45]{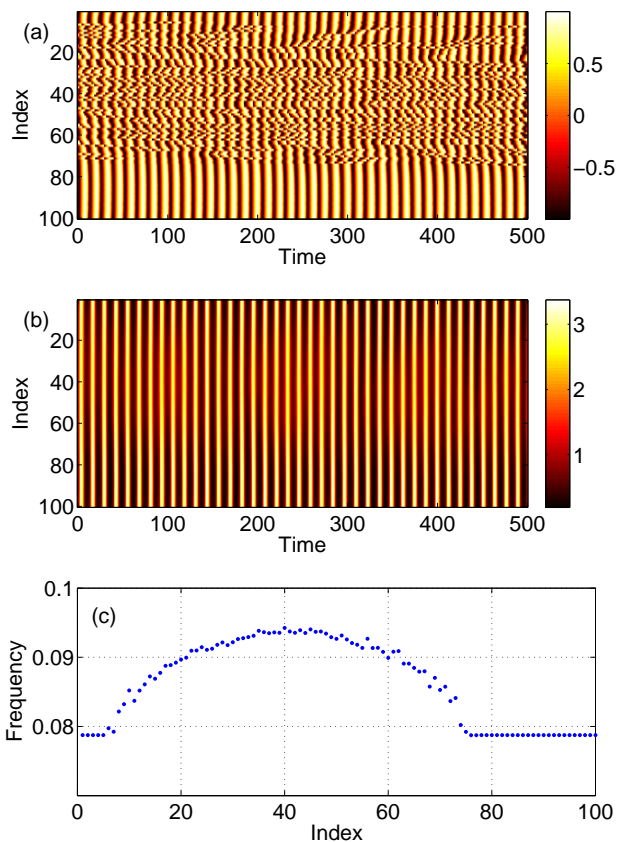}
\caption{(Color online) Chimera solution of the system~\eqref{eq:dwina}-\eqref{eq:dwinb}.
(a): $\sin{\theta_j}$; (b): $u_j$; (c): average rotation frequency of the $\theta_j$ over a simulation of duration $2000$
time units.
Parameters: $\omega_0=0.3,\sigma=0.001,n=4,L=4,\kappa=0.4,\beta=\pi/2-0.2,\epsilon=0.1,N=100$.}
\label{fig:win}
\end{figure}

A chimera state for~\eqref{eq:dwina}-\eqref{eq:dwinb} is shown in Fig.~\ref{fig:win}.
Setting $\epsilon=0$ in~\eqref{eq:dwinb} and solving for the $u_j$ one would obtain a nonlocally coupled
network of Winfree oscillators. Chimeras have been found in a network of two populations of Winfree
oscillators~\cite{pazmon13,lai14} but a truly nonlocally coupled network has apparently not yet been studied.

We have presented three one-dimensional networks, where each node is described by one variable which has a phase
associated with it and a second variable which is coupled in a diffusive fashion to just its two nearest
neighbours. All networks have the same structure and show chimera states over some range of parameters.
All have a small parameter ($\epsilon$) which controls the time scale of the diffusing variable,
so can be thought of as slow-fast systems~\cite{kue15}.

We have not given any stability analysis of the models presented here, only a self-consistency
argument for the first model. Chimeras in systems of the form~\eqref{eq:dthe0}
have been studied by passing to the continuum limit ($N\rightarrow\infty$) and analysing the resulting
continuity equation using the Ott/Antonsen ansatz~\cite{ottant09,ottant08,ome13,wolome11,lai09B}.
However, it does not seem that such an approach could be used to study the models presented here
due to the dynamics of the extra variables.

Regarding experimental implementation, note that the nonlocal coupling in the experiments reported
in~\cite{nkotin13,hagmur12} was implemented by computer, i.e.~the experiments were hybrid physical/computer.
The models presented here --- while being caricatures of physical systems --- 
have only local, nearest-neighbour diffusive-like coupling. Since diffusion is ubiquitous in
spatially extended systems of reacting chemicals, the most natural system in which to implement networks
of the form discussed here (without a computer) may be in
arrays of microsopic chemical oscillators~\cite{lidel14,toigon10,tintay10}.

%\bibliographystyle{plain}
%\bibliography{spiral}

\end{document}